\documentclass[twocolumn,showpacs,superscriptaddress,nofootinbib]{revtex4}
\usepackage{graphicx}

\begin{document}
\def\Journal#1#2#3#4{{#1} {\bf #2}, #3 (#4)}
\title{\bf $\Delta \phi \Delta \eta$ Correlations  in Central Au+Au Collisions
at $\sqrt{s_{NN}} =$  200 GeV}
\affiliation{Argonne National Laboratory, Argonne, Illinois 60439}
\affiliation{University of Birmingham, Birmingham, United Kingdom}
\affiliation{Brookhaven National Laboratory, Upton, New York 11973}
\affiliation{California Institute of Technology, Pasadena, California 91125}
\affiliation{University of California, Berkeley, California 94720}
\affiliation{University of California, Davis, California 95616}
\affiliation{University of California, Los Angeles, California 90095}
\affiliation{Carnegie Mellon University, Pittsburgh, Pennsylvania 15213}
\affiliation{Creighton University, Omaha, Nebraska 68178}
\affiliation{Nuclear Physics Institute AS CR, 250 68 \v{R}e\v{z}/Prague, Czech Republic}
\affiliation{Laboratory for High Energy (JINR), Dubna, Russia}
\affiliation{Particle Physics Laboratory (JINR), Dubna, Russia}
\affiliation{University of Frankfurt, Frankfurt, Germany}
\affiliation{Institute of Physics, Bhubaneswar 751005, India}
\affiliation{Indian Institute of Technology, Mumbai, India}
\affiliation{Indiana University, Bloomington, Indiana 47408}
\affiliation{Institut de Recherches Subatomiques, Strasbourg, France}
\affiliation{University of Jammu, Jammu 180001, India}
\affiliation{Kent State University, Kent, Ohio 44242}
\affiliation{Institute of Modern Physics, Lanzhou, China}
\affiliation{Lawrence Berkeley National Laboratory, Berkeley, California 94720}
\affiliation{Massachusetts Institute of Technology, Cambridge, MA 02139-4307}
\affiliation{Max-Planck-Institut f\"ur Physik, Munich, Germany}
\affiliation{Michigan State University, East Lansing, Michigan 48824}
\affiliation{Moscow Engineering Physics Institute, Moscow Russia}
\affiliation{City College of New York, New York City, New York 10031}
\affiliation{NIKHEF and Utrecht University, Amsterdam, The Netherlands}
\affiliation{Ohio State University, Columbus, Ohio 43210}
\affiliation{Panjab University, Chandigarh 160014, India}
\affiliation{Pennsylvania State University, University Park, Pennsylvania 16802}
\affiliation{Institute of High Energy Physics, Protvino, Russia}
\affiliation{Purdue University, West Lafayette, Indiana 47907}
\affiliation{Pusan National University, Pusan, Republic of Korea}
\affiliation{University of Rajasthan, Jaipur 302004, India}
\affiliation{Rice University, Houston, Texas 77251}
\affiliation{Universidade de Sao Paulo, Sao Paulo, Brazil}
\affiliation{University of Science \& Technology of China, Hefei 230026, China}
\affiliation{Shanghai Institute of Applied Physics, Shanghai 201800, China}
\affiliation{SUBATECH, Nantes, France}
\affiliation{Texas A\&M University, College Station, Texas 77843}
\affiliation{University of Texas, Austin, Texas 78712}
\affiliation{Tsinghua University, Beijing 100084, China}
\affiliation{Valparaiso University, Valparaiso, Indiana 46383}
\affiliation{Variable Energy Cyclotron Centre, Kolkata 700064, India}
\affiliation{Warsaw University of Technology, Warsaw, Poland}
\affiliation{University of Washington, Seattle, Washington 98195}
\affiliation{Wayne State University, Detroit, Michigan 48201}
\affiliation{Institute of Particle Physics, CCNU (HZNU), Wuhan 430079, China}
\affiliation{Yale University, New Haven, Connecticut 06520}
\affiliation{University of Zagreb, Zagreb, HR-10002, Croatia}
                                                                                                                                          
\author{J.~Adams}\affiliation{University of Birmingham, Birmingham, United Kingdom}
\author{M.M.~Aggarwal}\affiliation{Panjab University, Chandigarh 160014, India}
\author{Z.~Ahammed}\affiliation{Variable Energy Cyclotron Centre, Kolkata 700064, India}
\author{J.~Amonett}\affiliation{Kent State University, Kent, Ohio 44242}
\author{B.D.~Anderson}\affiliation{Kent State University, Kent, Ohio 44242}
\author{M.~Anderson}\affiliation{University of California, Davis, California 95616}
\author{D.~Arkhipkin}\affiliation{Particle Physics Laboratory (JINR), Dubna, Russia}
\author{G.S.~Averichev}\affiliation{Laboratory for High Energy (JINR), Dubna, Russia}
\author{Y.~Bai}\affiliation{NIKHEF and Utrecht University, Amsterdam, The Netherlands}
\author{J.~Balewski}\affiliation{Indiana University, Bloomington, Indiana 47408}
% ...
\collaboration{STAR Collaboration}\noaffiliation
\date{\today}

\begin{abstract}
We report charged-particle pair correlation analyses in the space of 
$\Delta \phi$ (azimuth) and $\Delta \eta$ (pseudo-rapidity), for central Au + 
Au collisions at $\sqrt{s_{NN}}$ = 200 GeV in the STAR detector. The analysis
involves unlike-sign charge pairs and like-sign charge pairs, which are 
transformed into charge-dependent (CD) signals and charge-independent (CI) 
signals. We present detailed parameterizations of the data. A model 
featuring dense gluonic hot spots as first proposed by Van Hove predicts
that the observables under investigation would have sensitivity to such a 
substructure should it occur, and the model also motivates selection of 
transverse momenta in the range $0.8 < p_t < 2.0$ GeV/c. Both CD and CI 
correlations of high statistical significance are observed, and possible
interpretations are discussed.
\end{abstract}


\maketitle
\section{introduction}  
The Search for a Quark-Gluon Plasma (QGP) \cite{whitepaper, QuarkM}
has been a high priority task at the Relativistic Heavy Ion Collider, RHIC 
\cite{RHIC}. Central Au + Au collisions at RHIC exceed \cite{transenergy} the 
initial energy density that is predicted by lattice Quantum Chromodynamics 
(QCD) to be sufficient for production of QGP \cite{LQCD}. Van Hove and others 
\cite{VanHove,LL,themodel} have proposed that bubbles localized in phase space 
(dense gluon-dominated hot spots) could be the sources of the final state 
hadrons from a QGP. Such structures would have smaller spatial dimensions than 
the region of the fireball viewed in this mid rapidity experiment and the 
correlations resulting from these smaller structures might persist in the 
final state of the collision. The bubble hypothesis has motivated this study 
and the model described in Ref. \cite{themodel} has led to our selection of
transverse momenta in the range $0.8 < p_t < 2.0$ GeV/c. The Hanbury-Brown and 
Twiss (HBT) results demonstrate that for $\sqrt{s_{NN}}$ = 200 GeV mid rapidity
central Au + Au, when $p_t >$ 0.8 GeV/c the average final state space geometry 
 for pairs close in momentum is approximately describable by dimensions of 
around 2 fm \cite{HBT}. This should lead to observable modification to the 
$\Delta \eta$ $\Delta \phi$ correlation. The present experimental analysis is 
model-independent and it probes correlations that could have a range of 
explanations.   

We present an analysis of charged particle pair correlations in two
dimensions --- $\Delta \phi$ and $\Delta \eta$ ---  based on 2 million 
central Au + Au events observed in the STAR detector at $\sqrt{s_{NN}}$ = 200 
GeV \cite{STARTPC}\footnote{$\Delta \phi = \phi_1 - \phi_2$ and 
$\Delta \eta = \eta_1 - \eta_2$}. The analysis leads to a multi-term 
correlation function (Section III C and E) which fits the 
$\Delta \eta \Delta \phi$ distribution well. It includes terms describing 
correlations known to be present: collective flow, resonance decays, and 
momentum and charge conservation. Data cuts are applied to make track merging 
effects, HBT correlations, and Coulomb effects negligible. Instrumental effects
resulting from detector characteristics are accounted for in the correlation 
function.  What remains are correlations whose origins are as yet unclear, and 
these are the main topic of this paper. We present high statistical precision 
correlations which can provide a quantitative test of the bubble model 
\cite{themodel} and other quantitative substructure models which may be 
developed. We also address possible jet phenomena. These precision data could 
stimulate other new physics ideas as possible explanations of the observed 
correlations. 

This paper is organized as follows. Section II describes the data utilized and 
its analysis. Section III describes finding a parameter set which fits the 
data well. Section IV presents and discusses the charge dependent (CD) and 
charge independent (CI) signals. Section V presents and discusses net 
charge fluctuation suppression. Section VI discusses the systematic errors.
Section VII is a discussion section. Section VIII contains the summary and 
conclusions. Appendices contain explanatory details.

\section{data analysis}
\subsection{ Data Utilized}
The data reported here is the full sample of STAR events taken at RHIC 
during the 2001 running period for central Au + Au collisions at 
$\sqrt{s_{NN}}$ = 200 GeV. The data were taken using a central trigger
with the full STAR magnetic field (0.5 Tesla).

The central trigger requires a small signal, coincident in time, in each of 
two Zero Degree Calorimeters, which are positioned so as to intercept spectator
neutrons, and also requires a large number of counters in the Central Trigger 
Barrel to fire. Approximately 90\% of the events are in the top 10\% of the
minimum bias multiplicity distribution, which is called the 0-10\% centrality 
region. About 5\% are in the 10-12\% centrality region with the remainder 
mostly in the 12-15\% centrality region. To investigate the sensitivity of our 
analyses to the centrality of the data sample fitted, we compared fits of the 
entire data sample to fits of the data in the 5-10\% centrality region. Both 
sets of fits had consistent signals within errors. Thus no significant 
sensitivity to the centrality was observed in the correlation data from the 
central triggers.

\begin{figure*}[ht] \centerline{\includegraphics[width=0.800\textwidth]
{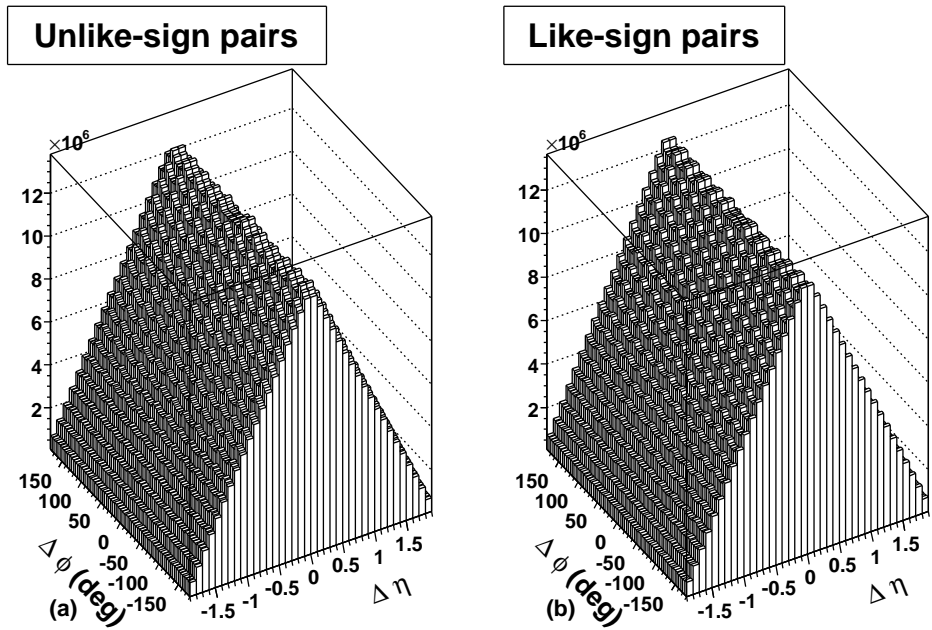}} \caption[]{a) left side: The $\Delta \phi$ - $\Delta \eta$ 
correlation data for unlike-sign charge particle pairs from the Star central 
trigger dataset shown in a 2-dimensional (2-D) perspective plot. The particle
tracks have $0.8$ GeV/c $ < p_t < 2.0$ GeV/c and $|\eta| < 1.0$. The structure 
that looks like tiles on a roof is due to the readout boundary effects of the 
12 sector TPC. 

b) right side:  The similar correlation data
for like-sign charge particle pairs is shown.} 
\label{figure1}
\end{figure*}

About half the data were taken with the magnetic field parallel to the beam
axis direction (z) and the other half in the reverse field direction in 
order to determine if directional biases are present. As discussed later in 
this subsection, our $\chi^2$ analyses demonstrated there was no evidence of 
any difference in the data samples from the two field directions, and thus no
evidence for directional biases. 

The track reconstruction for each field direction was done using the same 
reconstruction program. Tracks were required to have at least 23 hits 
in the TPC (which for STAR eliminates split tracks), and to have 
pseudo-rapidity, $\eta$, between -1 and 1. Each event was required to have at 
least 100 primary tracks. These are tracks that are consistent with the 
criteria that they are produced by a Au + Au beam-beam interaction. This cut 
rarely removed events. The surviving events totaled 833,000 for the forward 
field and 1.1 million for the reverse field. The transverse momentum selection 
$0.8 < p_t  < 2.0$ GeV/c was then applied. 

Based on the z (beam axis) position of the primary vertex the events were 
sorted into ten 5 cm wide bins covering -25 cm to +25 cm.  The events for the 
same $z$ bin, thus the same acceptance, were then merged to produce 20 files, 
one for each z bin for each sign of the magnetic field.

The files were analyzed in two-dimensional (2-D) histograms of the difference 
in $\eta$ ($\Delta \eta$), and the difference in $\phi$ ($\Delta  \phi$) for 
all the track pairs in each event. Each 2-D histogram had 72 $\Delta \phi$ 
bins ($5^\circ$) from $-180^\circ$ to $180^\circ$ and 38 $\Delta \eta$ bins 
(0.1) from -1.9 to 1.9. The sign of the difference variable was chosen by 
labeling the positive charged track as the first of the pair for the 
unlike-sign charge pairs, and the larger $p_t$ track as the first for the 
like-sign charge pairs. Our labeling of the order of the tracks in a pair 
allows us to range over four $\Delta \phi$ - $\Delta \eta$ quadrants, and 
investigate possible asymmetric systematic errors due to space, magnetic field 
direction, behavior of opposite charge tracks, and systematic errors dependent 
on $p_t$. Our consistently satisfactory results for our extensive $\chi^2$ 
tests of these quadrants for fits to this precision data revealed no evidence 
for such effects.

Then we compared the $\Delta \phi$ - $\Delta \eta$ data for the two field
directions on a bin by bin basis. In the reverse field data, we reversed the
track curvature due to the change in the field direction, and changed the sign 
of the z axis making the magnetic field be in the same direction as the 
positive z direction. This is done by reflecting along the z axis, and 
simultaneously reflecting along the y axis. In the two dimensional 
$\Delta \phi$ - $\Delta \eta$ space this transformation is equivalent to a 
reflection in $\Delta \phi$ and $\Delta \eta$. For each pair we changed the 
sign of its $\Delta \phi$ and $\Delta \eta$ in the reverse field data. We then 
calculated a $\chi^2$ based on the difference between the forward field and the
reverse field, summing over the $\Delta \phi$ - $\Delta \eta$ histograms 
divided by the errors added in quadrature. The resulting $\chi^2$ for the two 
fields showed an agreement to within $1.5\sigma$. Therefore we added the data 
for the two field directions.

We also found there was no significant dependence within $2.2\sigma$ on the 
vertex z co-ordinate. Following the same methodology we added the files for 
those 10 bins also.

\begin{figure*}[ht] \centerline{\includegraphics[width=0.800\textwidth]
{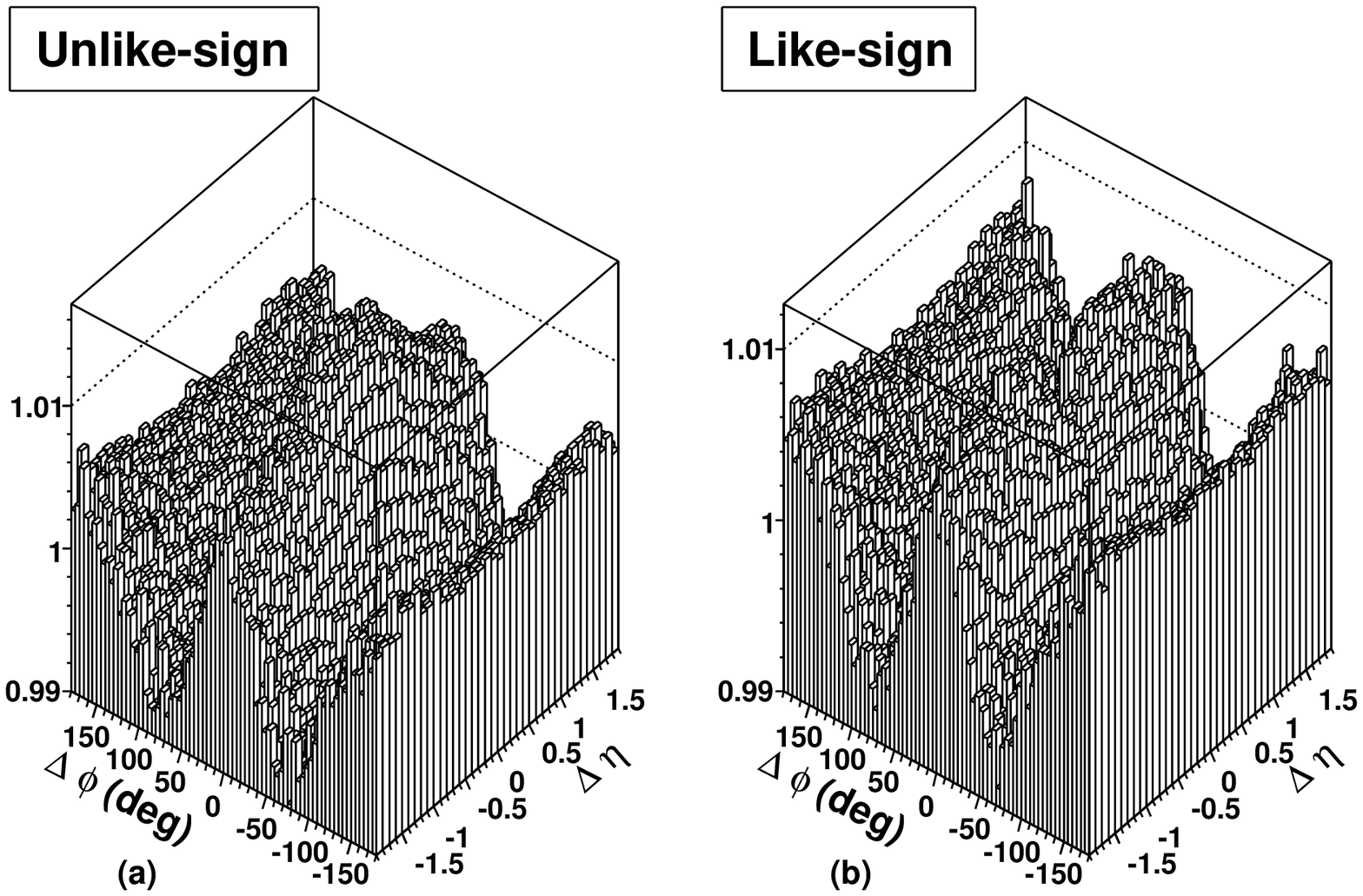}} \caption[]{a) left side: The correlation data for the ratio of 
the histograms of same-event-pairs to mixed-event-pairs for unlike-sign charged
pairs, shown in a two-dimensional (2-D) perspective plot $\Delta \phi$ - 
$\Delta \eta$. The plot was normalized to a mean of 1.

b) right side: The similar correlation data for like-sign charge pairs.
 }\label{figure2}
\end{figure*}

\subsection{Analysis Method}
Separate $\Delta \phi$ - $\Delta \eta$ histograms were made for unlike-sign
charge pairs and like-sign charge pairs from the same-event-pairs, since their 
characteristics were different. Both histograms are needed to later determine 
the CD and the CI correlations as defined in Sections IV A and B. These data 
are shown in Fig. 1. Similar histograms were made with each track paired with
tracks from a different event (mixed-event-pairs), adjacent in time, from the 
same z vertex bin. This allows the usual technique of dividing the histograms
of the same-events-pairs by the histograms of the mixed-events-pairs which 
strongly suppresses instrumental effects such as acceptance etc., but leaves 
small residual effects. These include those due to time dependent efficiency of
the tracking in the readout boundary regions between the 12 TPC sectors, which 
had small variations in space charge from event to event. Also space charge in 
general can cause track distortion and efficiency variation event by event. 
The ratio of same-event-pairs to mixed-event-pairs histograms is shown in Fig.
2 where the plot was normalized to a mean of 1.

The expected symmetries in the data existed which allowed us to fold all four 
$\Delta \phi$ - $\Delta \eta$ quadrants into the one quadrant where both 
$\Delta \phi$ and $\Delta \eta$ were positive. After the cuts described in 
subsection C, we compared the unfolded bins to the folded average for 
unlike-sign charge pairs and like-sign charge pairs separately. The 
$\chi^2$ were less than the Degrees of Freedom (DOF), and the $\chi^2$/DOF
were within 2-3$\sigma$ of 1. 

Thus though we searched carefully in a number of ways to find asymmetries in 
the data via extensive $\chi^2$ analyses and observation of fit behavior, none 
of any significance were found. By folding four quadrants into one, we 
quadrupled the statistics in each bin analyzed. Fig. 3 shows this folded data 
after cuts (see subsection C) were made to make track merging, Coulomb, and 
HBT effects negligible. Henceforth the folded and cut data will be used for 
our fits.

\begin{figure*}[ht] \centerline{\includegraphics[width=0.800\textwidth]
{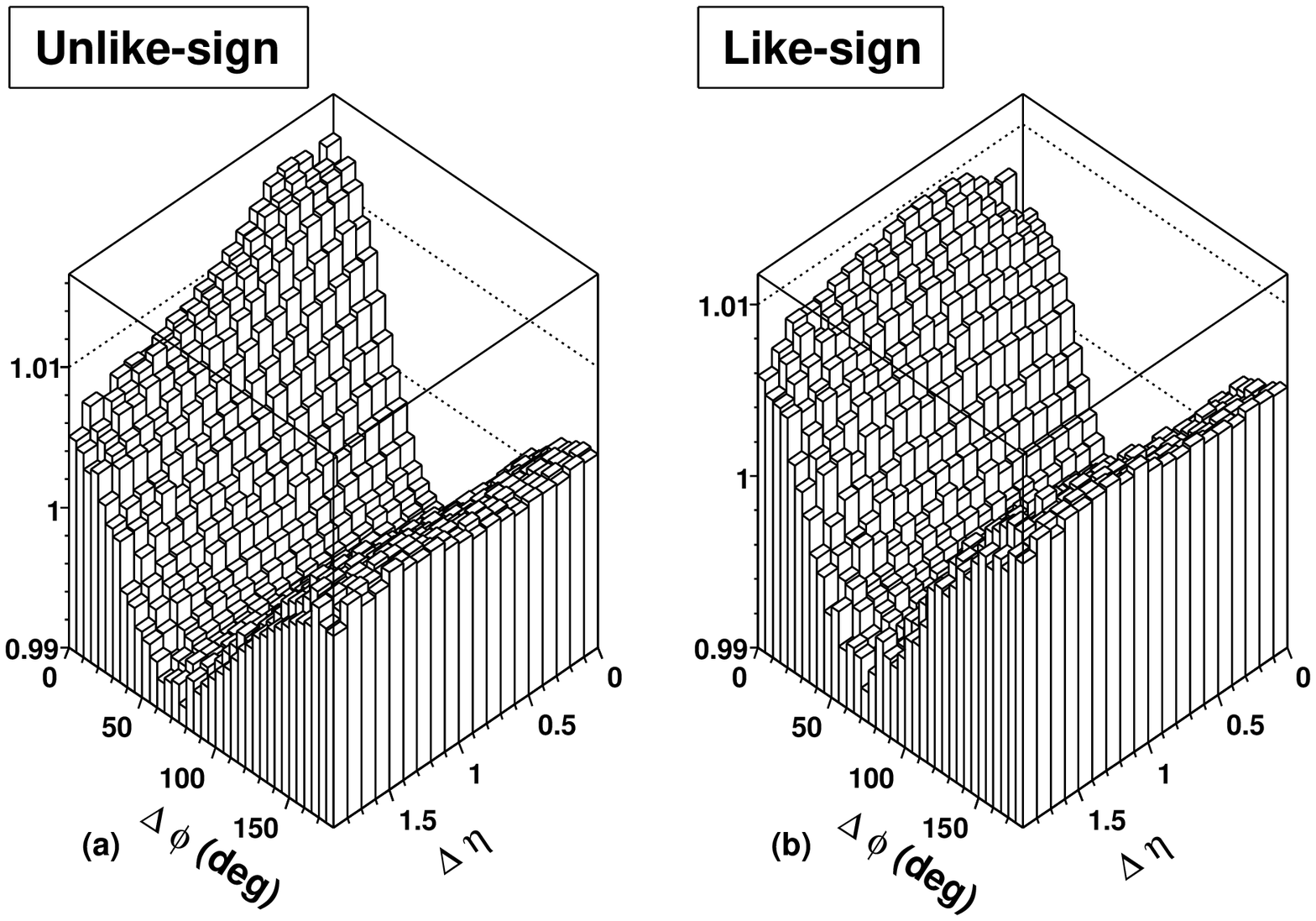}} \caption[]{a) left side: Folded correlation data for unlike-sign 
charge pairs on $\Delta \phi$ and $\Delta \eta$ based on demonstrated 
symmetries in the data. This increases the statistics in a typical bin by a 
factor of four. Henceforth we will be dealing with folded data only.
b) right side: Similar folded data for like-sign charge pairs.
} \label{figure3}
\end{figure*}

\subsection{Cuts}
At small $\Delta \phi$ and small $\Delta \eta$ (i.e. small space angles) track
merging effects occur. To determine the cuts needed to reduce these effects to 
a negligible level, we varied small $\Delta \phi$ and small $\Delta \eta$ cuts.
Simultaneously the $\chi^2$ of an approximate fit to the data using equation 
(3) + (4) + (5) was studied as a function of the bins included in the fit. 
With larger cuts the $\chi^2$ behaved properly until one or more of the bins 
cut out was included in the fit. This caused a huge increase in $\chi^2$, 
revealing that those bin(s) were distorted by merging effects. We confirmed by 
visual inspection that track merging effects clearly became important in the 
bins cut-out. The track merging caused a substantial reduction in track 
recognition efficiency, and supported the quantitative results of our $\chi^2$ 
analysis. The sharp sensitivity of the $\chi^2$ to the cuts showed how 
localized the merging losses were. We adopted these cuts and they are included 
in the folded data analyzed

For the unlike-sign charge pairs we cut out the data from the regions
0.0 $< \Delta \eta <$ 0.1 when $0^\circ < \Delta \phi < 20^\circ$,
and 0.1 $< \Delta \eta <$ 0.2 when $0^\circ < \Delta \phi < 10^\circ$.
These cuts to make track merging effects negligible also made the Coulomb 
effect negligible, since it is effective within a $\Delta$p of 30 MeV$/c$. 
This corresponds to a $2^\circ$ opening angle at $p_t$ of 0.8 GeV/c. The 
$2^\circ$ along the beam axis near $90^\circ$ corresponds to a $\Delta \eta$ 
of 0.035. 

For the like-sign charge pairs we cut out the data from the region
0.0 $< \Delta \eta <$ 0.2 when $0^\circ < \Delta \phi < 5^\circ$.
The HBT effect is already small because of track merging within a 
$\Delta$p of 50 MeV/c, corresponding to a $4^\circ$ opening angle at 0.8 
GeV/c. The $4^\circ$ along the beam corresponds to a $\Delta \eta$ of 0.07. 
All the cuts described above were applied to the mixed events also. Thus the 
final track merging cuts selected reduce the tracking problems due to overlap 
and merging to negligible levels, and also reduce Coulomb and HBT effects to 
negligible levels.

The fits to the subsequently shown data were made over the whole 
$\Delta \phi$ range. The data for $|\Delta \eta| > 1.5$
were cut out, since the statistics were 
low (see Fig. 1) and tracking efficiency varies at higher $|\Delta \eta|$.

\section{Parameterization of data.} 
We want to obtain a set of functions which will fit the data well, and are 
interpretable, to the extent practical. We utilized parameterizations
representing known, expected physics, or attributable to instrumentation 
effects. Any remaining terms required to obtain good fits to the data can be 
considered as signals of new physical effects. Thus signal $\equiv$ data - 
(known and expected) effects. The three known effects: elliptic flow,
residual instrumental effects, momentum and charge conservation terms were
parameterized. We then find parameters for the signal terms which are necessary
in order to achieve a good fit to our high statistical precision data.

In the following we discuss the relative importance of the effects which are
determined by $\chi^2$ in our final overall fit to equation (3) + (4) + (5). 

\subsection{Elliptic Flow}
Elliptic flow is a significant contributor to two-particle correlations in
heavy ion collisions, and must be accounted for in our analysis. Elliptic
flow is not a significant contributor to two-particle correlations in  p p
collisions. The elliptic flow for STAR Au + Au data has been investigated in an
extensive series of measurements and analyses \cite{flow} using four 
particle cumulant methods and two particle correlations. This determines a 
range of amplitudes($v_2$) for $\cos$(2$\Delta \phi$) terms between the four 
particle cumulant which is the lower flow range boundary of $v_2$($v_2$=0.035),
and the average value of the four and two particle cumulant which is the upper
range boundary of $v_2$($v_2$=0.047). These results were determined for the 
$p_t$ range 0.8 to 2 GeV/c by weighting the central trigger spectrum as a 
function of multiplicity and centrality compared to the minimum bias trigger 
data for which the $v_2$ was measured.

The constraint is used in the following way. We find our best fit to the data 
by allowing the parameters in the fit to vary freely until the $\chi^2$ is 
minimized. However, there is in the case of the $v_2$ parameter a constraint 
that it must lie in the flow range cited above. Since our best fit was 
consistent with the bottom of the flow range it was not affected by the 
constraint. The value of $v_2$ cannot be varied to lower values because of the 
constraint and thus our best fit corresponds to the lower boundary of the flow
range and had a $v_2$ of 0.035. However, $v_2$ can be varied to higher values 
until either we reach the upper boundary or the fit worsens by $1\sigma$ 
which is a $\chi^2$ change of 32. As discussed in Section VI, our fit worsens 
by $1\sigma$ when $v_2$ is increased to 0.042 well within the flow range 0.035
to 0.047. This represents a systematic error due to flow and is the dominant 
error( see Section VI ). In Appendix A, a detailed description of 
multi-parameter fitting and error range determinations are given.  

\subsection{Instrumental Effects}
In addition to the tracking losses resulting from overlapping tracks in 
the TPC, which we handled by cuts, there are tracks lost in the boundary
areas of the TPC between the 12 sectors. These regions result in a loss of 
acceptance in the $\phi$ measurements since the particle tracks cannot be
readout by electronics. This produces a $\phi$ dependence with a period of 
$30^\circ$ which is greatly reduced from about a 4\% amplitude to about 0.02\% 
(reduction by a factor of 200), by the normalization to mixed-event-pairs. 
However it is not completely eliminated, since STAR still has small time 
dependent variations from event to event, such as space charge effects and 
slight differences in detector and beam behavior. To correct for the residual 
readout  $\Delta \phi$ dependence, a term with a $30^\circ$ period to represent
the TPC boundary periodicity and a first harmonic term with a $60^\circ$ period
were used. The unlike-sign and like-sign charge pairs behave differently over 
the TPC boundary regions in $\phi$, because positive tracks are rotated in one 
direction and negative tracks are rotated in the opposite direction. This 
requires that the sector terms have an independent phase associated with each. 
In the Tables I and II the terms are labeled ``sector'' and ``sector2'', 
``phase'' and ``phase2''.  The functional form for these sector effects is:

\begin{equation}
\rm Sectors \it (\Delta \phi) = S \sin(12\Delta \phi - \varphi) + 
S_2 \sin(6\Delta \phi - \varphi_2) 
\end{equation}

There is a $\phi$ independent effect which we attribute to losses in the larger
$\eta$ tracking in the TPC. We utilized mixed-event-pairs with a similar 
$z$-vertex to take into account these losses. Imperfections in this procedure 
leave a small bump near $\Delta \eta$ = 1.15 to be represented in the fit by 
the terms labeled etabump amp and etabump width in the tables. The width of 
this bump should be independent of the charge of the tracks, so we constrained 
it to be the same for like and unlike-sign charge pairs to improve the fit
stability. Thus the functional form for this instrumental effect is:
 
\begin{equation} \rm Etabump \it (\Delta \eta) = E e^{-(\Delta \eta - 1.15)^2/2\sigma_E^2} \end{equation} 
                           
In our final $3\sigma$ fits for the combined like and unlike-sign charge pairs 
(See Tables I and II), leaving out the instrumental corrections would cause 
our $3\sigma$ fit to deteriorate to a $16\sigma$ fit which is unacceptable. 
It should be noted that standard $\chi^2$ analyses are often not considered 
credible if the fit exceeds 2-3$\sigma$ statistical significance. 

\subsection{Correlations associated with Momentum and Charge Conservation}
It is important to ensure that momentum and charge conservation correlation 
requirements are satisfied. For random emission of single particles with 
transverse momentum conservation globally imposed, a negative 
$\cos$($\Delta \phi$) term alone can represent this effect. However, the 
complex correlations that occur at RHIC result in multiple uncorrelated sources
which are presently not understood. It was not possible to fit our data with 
the $\cos$($\Delta \phi$) term alone. This fit was rejected by $100\sigma$. 
This was not surprising since random emission of single particles with momentum
conservation would not lead to the particle correlations observed at RHIC.
Therefore we suspected that a more complete description of the momentum and
charge conservation was required. No one has succeeded in solving this complex
problem in closed form even in the theoretical case where you observe all
particles. Hence a reasonable approach was to try to solve it for the tracks we
are observing in order to obtain a good fit. We used Fourier expansion in the 
two variables we have, namely $\Delta \phi$ and $\Delta \eta$.

Assuming that the $\cos$($\Delta \phi$) term for random single particle 
emission was the first term in a Fourier expansion of odd terms, a second term 
$\cos$(3$\Delta \phi$) was added and found to account for about 95\% of the 
$100\sigma$ rejection. Based on the residual analysis we concluded the 
remaining 5\% required $\Delta \eta$ dependent terms for its removal. Therefore
we multiplied terms of the type $\cos$($\Delta \phi$) and 
$\cos$(3$\Delta \phi$) by a $\Delta \eta$ dependent polynomial expansion 
cutting off at $\Delta \eta^2$. This essentially removed the remaining 5\%
rejection.

Some of the $\Delta \eta$ dependent terms make quite small contributions in the
fits, but in the aggregate they improve our overall fit by about $4\sigma$. 
Thus a good fit of $3\sigma$ is downgraded to an unacceptable one of $7\sigma$ 
without the $\Delta \eta$ dependent expansion, since they add. One should note 
that fits which exceed $3\sigma$ are often not considered credible by 
experienced practitioners of precision data analysis, and a precision data 
analysis was a prime objection of this paper. In addition we found that we 
needed an overall $\Delta \eta^2$ dependent fit parameter.

If we take the sum of the terms we have in the above subsections A, B and C
 we obtain 

 \begin{eqnarray}
\bf Bk \it = \rm (Known \it + \rm Expected)Effects \it = \nonumber \\
B_{00} + B_{02}\Delta \eta^2 + B_{10}\cos\Delta \phi + B_{11}\Delta \eta\cos\Delta \phi \nonumber \\
+ B_{12}\Delta \eta^2\cos\Delta \phi 
 + 2 v_2^2\cos (2 \Delta \phi) + B_{30}\cos(3\Delta \phi) \nonumber \\
+ B_{31}\Delta \eta\cos(3\Delta \phi) +B_{32}\Delta \eta^2\cos(3\Delta \phi) \nonumber \\
+ \rm Sectors \it (\Delta \phi) + \rm Etabump \it (\Delta \eta) \nonumber \\
        \end{eqnarray}

\begin{center}
\begin{table}
\begin{tabular}{ c r }\hline\hline
Name  & Value \\ \hline
lump amplitude ($A_u$) &  0.02426$^{-0.00498}_{+0.00048}$\\ \hline
lump $\Delta\phi$ width ($\sigma_{\Delta\phi}$) & 32.68$^{-2.12}_{+0.79}$\\ \hline
lump $\Delta\eta$ width ($\sigma_{\Delta\eta}$) & 1.058$^{-0.137}_{+0.065}$\\ \hline
fourth   (f)     &  0.100$^{+0.031}_{-0.028}$\\ \hline
 \hline
constant   ($B_{00}$)  &  0.99497$^{+0.00142}_{-0.00021}$\\ \hline
$\Delta\eta^2$   ($B_{02}$) &  0.00078$\mp 0.00018$\\ \hline
$\cos\Delta\phi$   ($B_{10}$) & -0.00710$^{+0.00230}_{-0.00035}$\\ \hline
$\Delta\eta\cos\Delta\phi$  ($B_{11}$)  &  0.00092$\mp 0.00078$\\ \hline
$\Delta\eta^2\cos\Delta\phi$  ($B_{12}$) & -0.00049$\mp 0.00057$\\ \hline
$\cos 3\Delta\phi$   ($B_{30}$)    &  0.00058$^{+0.00028}_{-0.00020}$\\ \hline
$\Delta\eta\cos 3\Delta\phi$ ($B_{31}$)  &  0.00030$\pm 0.00062$\\ \hline
$\Delta\eta^2\cos 3\Delta\phi$ ($B_{32}$) & -0.00014$\mp 0.00044$\\ \hline
\hline
sector  (S)        &  0.00016$\pm 0.00006$\\ \hline
phase  ($\varphi$)       &  8.6$\pm 2.1$\\ \hline
sector2   ($S_2$)   &  0.00006$\pm 0.00007$\\ \hline
phase2  ($\varphi_2$) &  23.0$\pm 10.0$\\ \hline
$\Delta\eta$bump amp (E) &  0.00030$\mp 0.00025$\\ \hline
$\Delta\eta$bump width ($\sigma_E$) &  0.189 (fixed) \\ \hline
\hline
$\chi^2/DOF$       &  572 / 517         \\ \hline\hline
\end{tabular}
\caption{Unlike-Sign Charge Pairs Fit Parameters for equation (3) + 
equation (4). The table has 3 sections. The top section lists parameter names, 
values and errors for the Approximate Gaussian Signal fit (lump). Fourth(f) is 
the additional term in the exponent. The source of the above is Subsection III 
E. The upper error is the change in each parameter when one increases the 
elliptic flow until the fit is $1\sigma$ worse. The lower error is determined 
by varying each parameter at the lower range of the elliptic flow while all the
other parameters are free to readjust until the fit loses $1\sigma$ in 
significance. The middle section has the normalization and a small background 
term followed by six Momentum and Charge Conservation terms. Source is 
Subsection III C. The bottom section has instrumental terms. Four terms which 
are due to TPC sector gaps and two due to large $\eta$ tracking errors. 
Source is Subsection III B. Uncertainties are dominantly systematic, assessed 
as described in Appendix A.}
\end{table}
\end{center}

\begin{center}
\begin{table}
\begin{tabular}{ c r }\hline\hline
Name & Value \\ \hline
lump amplitude ($A_l$)  & 0.01823$^{-0.00482}_{+0.00069}$\\ \hline
lump $\Delta\phi$ width ($\sigma_{\Delta\phi l}$) &   32.02$^{-2.91}_{+1.02}$\\ \hline
lump $\Delta\eta$ width ($\sigma_{\Delta\eta l}$) &   1.847$^{-0.315}_{+0.220}$ \\ \hline
dip amplitude ($A_d$)   &-0.00451$^{-0.00092}_{+0.00090}$ \\ \hline
dip $\Delta\phi$ width ($\sigma_{\Delta\phi d}$) &   14.23$^{-2.64}_{+2.91}$\\ \hline
dip $\Delta\eta$ width  ($\sigma_{\Delta\eta d}$) &  0.228$^{+0.050}_{-0.041}$\\ \hline
\hline
constant    ($B_{00}$) &  0.99581$^{+0.00136}_{-0.00019}$\\ \hline
$\Delta\eta^2$    ($B_{02}$)&  0.00100$\mp .00017$\\ \hline
$\cos\Delta\phi$   ($B_{10}$) & -0.00737$^{-0.00221}_{-0.00040}$\\ \hline
$\Delta\eta\cos\Delta\phi$ ($B_{11}$)   &  0.00075$^{+0.00072}_{-0.00073}$\\ \hline
$\Delta\eta^2\cos\Delta\phi$ ($B_{12}$) & -0.00015$^{-0.00054}_{+0.00055}$\\ \hline
$\cos 3\Delta\phi$   ($B_{30}$)  &  0.00070$^{+0.00033}_{-0.00020}$\\ \hline
$\Delta\eta\cos 3\Delta\phi$  ($B_{31}$)  & -0.00027$^{-0.00064}_{+0.00065}$\\ \hline
$\Delta\eta^2\cos 3\Delta\phi$ ($B_{32}$) &  0.00026$^{+0.00044}_{-0.00045}$\\ \hline
\hline
sector (S)    &  0.00021$\pm 0.00006$\\ \hline
phase ($\varphi$) &   22.6$\mp 1.5$\\ \hline
sector2  ($S_2$)     &  0.00007$\pm 0.00007$\\ \hline
phase2  ($\varphi_2$)  &  32.8$\mp 7.0$\\ \hline
$\Delta\eta$bump amp (E) &  0.00022$^{+0.00024}_{-0.00022}$\\ \hline
$\Delta\eta$bump width ($\sigma_E$)&  0.189 (fixed)  \\ \hline
\hline
$\chi^2/DOF$      &  588 / 519         \\ \hline\hline
\end{tabular}
\caption{Like-Sign Charge Pairs Fit Parameters for equation (3) + 
equation (4). The table has 3 sections. The top section lists parameter names, 
values and errors for the Primary Gaussian Signal fit (lump). This is followed 
by a much smaller Gaussian (dip). Source is Subsection III E. The upper error 
is the change in each parameter when one increases the elliptic flow until the 
fit is $1\sigma$ worse. The lower error is determined by varying each parameter
at the lower range of the elliptic flow while all the other parameters are free
to readjust until the fit loses $1\sigma$ in significance. The middle section 
has the normalization and a small background term, followed by six Momentum and
Charge Conservation terms. Source is Subsection III C. The bottom section has 
instrumental terms. Four terms which are due to TPC sector gaps and two due to 
large $\eta$ tracking errors. Source is Subsection III B. Uncertainties are 
dominantly systematic, assessed as described in Appendix A.}
\end{table}
\end{center}

\subsection{Fitting with Bk}

 We used the well known result \cite{probability} that for a large number of
degrees of freedom (DOF), where the number of parameters is a small fraction of
DOF and the statistics are high, the $\chi^2$ distribution is normally
distributed about the DOF. The significance of the fit decreases by $1\sigma$
whenever the $\chi^2$ increase is equal to $\sqrt{2(DOF)}$ which for
our 517-519 DOF is equal to 32. Appendix A gives further details.

If we fit the functional form of \bf Bk \rm (equation (3)) to both the 
unlike-sign charge pairs and the like-sign charge pairs, i.e. the whole data
 set, $\chi^2$ is about 10,400 for about 
1045 degrees of freedom. 

The standard deviation on 1045 DOF is about 46 so the fit is rejected by 
around $200\sigma$. We used many free parameters (15) in equation (3) 
yet additional unknown terms appear to be so sharply varying that it is 
clearly impossible for the \bf Bk \rm functional form to fit the data. 
A more detailed discussion of the statistical methods used in the data 
analyses, and the treatment of systematic and other errors in this paper is 
given in Appendix A.

\subsection{Signal Terms}

Many signal terms in physics are gaussian-like. We therefore tried fitting the 
signal data using two dimensional (2-D) gaussians or approximate gaussians. The
unlike-sign charge pairs correlations were well fit ($2\sigma$), by adding to 
\bf Bk \rm an additional 2-D approximate gaussian in $\Delta \eta$ and 
$\Delta \phi$ (Fig. 4a) given by:

\begin{equation} \rm Unlike-sign\>Signal \it  =  A_u e^{-(\Delta \phi^2/2\sigma_\phi^2 +\Delta 
\eta^2/2\sigma_\eta^2 -f \Delta \eta^4)}\end{equation}

Considering the enormous improvement in fit quality afforded by the addition
of this signal term, we conclude that this function in  equation (4) provides a
compact analytic description of the signal component of the unlike-sign charged
pairs correlation data. The fit was improved by the addition of a term 
dependent on ($\Delta \eta^4$) in the exponent (called ``fourth'' in Table I). 
The fit for the unlike-sign charge pairs signal data is shown in Fig. 4a. 
The unlike-sign charge pairs signal data corresponding to the fit 
(unlike-sign charge pairs data minus \bf Bk \rm) is shown in Fig. 4b.

The like-sign charge pairs data which also could not be fit by \bf Bk \rm 
alone, was well described ($2\sigma$) when we added (see Fig. 3b) a positive 
2-D gaussian and a small negative 2-D gaussian dip given by: 
 
\begin{eqnarray} \rm Like-sign\>Signal \it  = A_l e^{-(\Delta \phi^2/2\sigma_{\phi l}^2 +\Delta
 \eta^2/2\sigma_{\eta l}^2)} \nonumber \\
+ A_d e^{-(\Delta \phi^2/2\sigma_{\phi d}^2 
+\Delta \eta^2/2\sigma_{\eta d}^2)}\end{eqnarray}

Therefore, we conclude that equation (5) provides an efficient description of
the signal component of the like-sign charge pairs data. The large signal 
referred to in Table II as ``lump'' in the like-sign charge pairs correlation 
is a 2-D gaussian centered at the origin. It is accompanied by a small narrower
2-D gaussian ``dip'' (also centered at the origin) subtracted from it. 
The terms are labeled in the fits as: lump amplitude, dip amplitude, lump 
$\Delta \phi$ width, lump $\Delta \eta$ width, dip $\Delta \phi$ width, and 
dip $\Delta \eta$ width. Note that the volume of the dip is about 1.6\% of the 
volume of the large lump signal. However, if we neglected to include this small
dip, our fit deteriorates by $28\sigma$. This dip might be caused by 
suppression of like-sign charge particle pair emission from localized neutral 
sources such as gluons. The function fit to the like-sign charged pairs signal 
data is shown in Fig. 5a. The like-sign charge pairs signal data 
(like-sign charge pairs data minus \bf Bk) \rm is shown in Fig. 5b. The 
uncertainties quoted throughout this paper corresponds to a change in $\chi^2$
of $\Delta\chi^2=32$, rather than the more commonly used $\Delta\chi^2=1$
(see Appendix A for more details).

\subsection{Summary of Parameterizations}
Equation (3) + (4) yields a $2\sigma$ fit for the unlike-sign charge pairs 
correlation, and thus is an appropriate and sufficient parameter set.

Equation (3) + (5) yields a $2\sigma$ fit for the like-sign charge pairs 
correlation, and thus is an appropriate and sufficient parameter set. 

Equation (3) + (4) + (5) yields a $3\sigma$ fit for the entire data set,
unlike-sign charge pairs correlation + like-sign charge pairs correlation, and
thus is an appropriate and sufficient parameter set for the entire data set.

\begin{figure*}[ht] \centerline{\includegraphics[width=0.800\textwidth]
{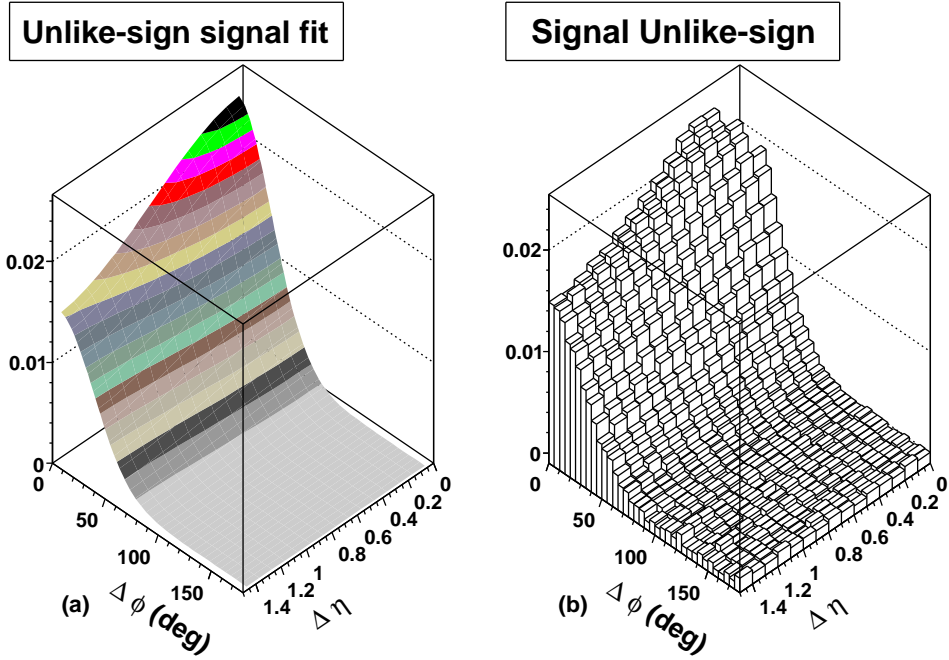}} \caption[]{``(Color online)''a) left side: The 2 dimensional 
(2-D) approximately gaussian signal shape equation (4) from the fit to the 
normalized and folded unlike-sign charge pairs signal data.

b) right side: The unlike-sign charge pairs signal data corresponding to the 
adjoining signal function on the left side.}
\label{figure4}
\end{figure*}

\begin{figure*}[ht] \centerline{\includegraphics[width=0.800\textwidth]
{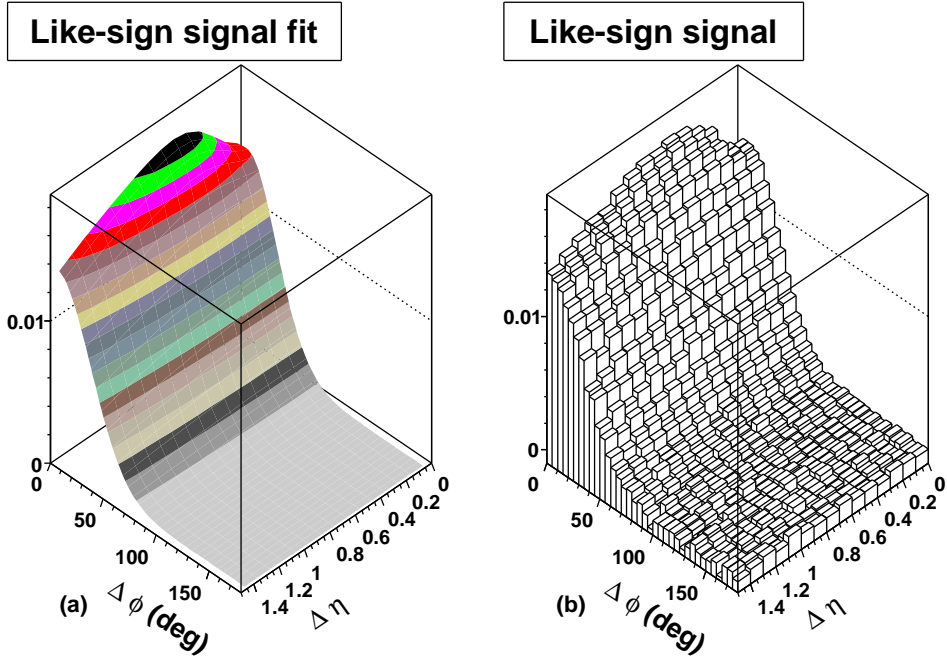}} \caption[]{``(Color online)''a) left side: A  perspective plot 
of the fit to the 2-D like-sign charge pairs signal shape of equation (5).

b) right side:  Like-sign charge pairs signal data corresponding to the 
adjacent signal fit on the left side.} 
\label{figure5}
\end{figure*}
 
\begin{figure*}[ht] \centerline{\includegraphics[width=0.800\textwidth]
{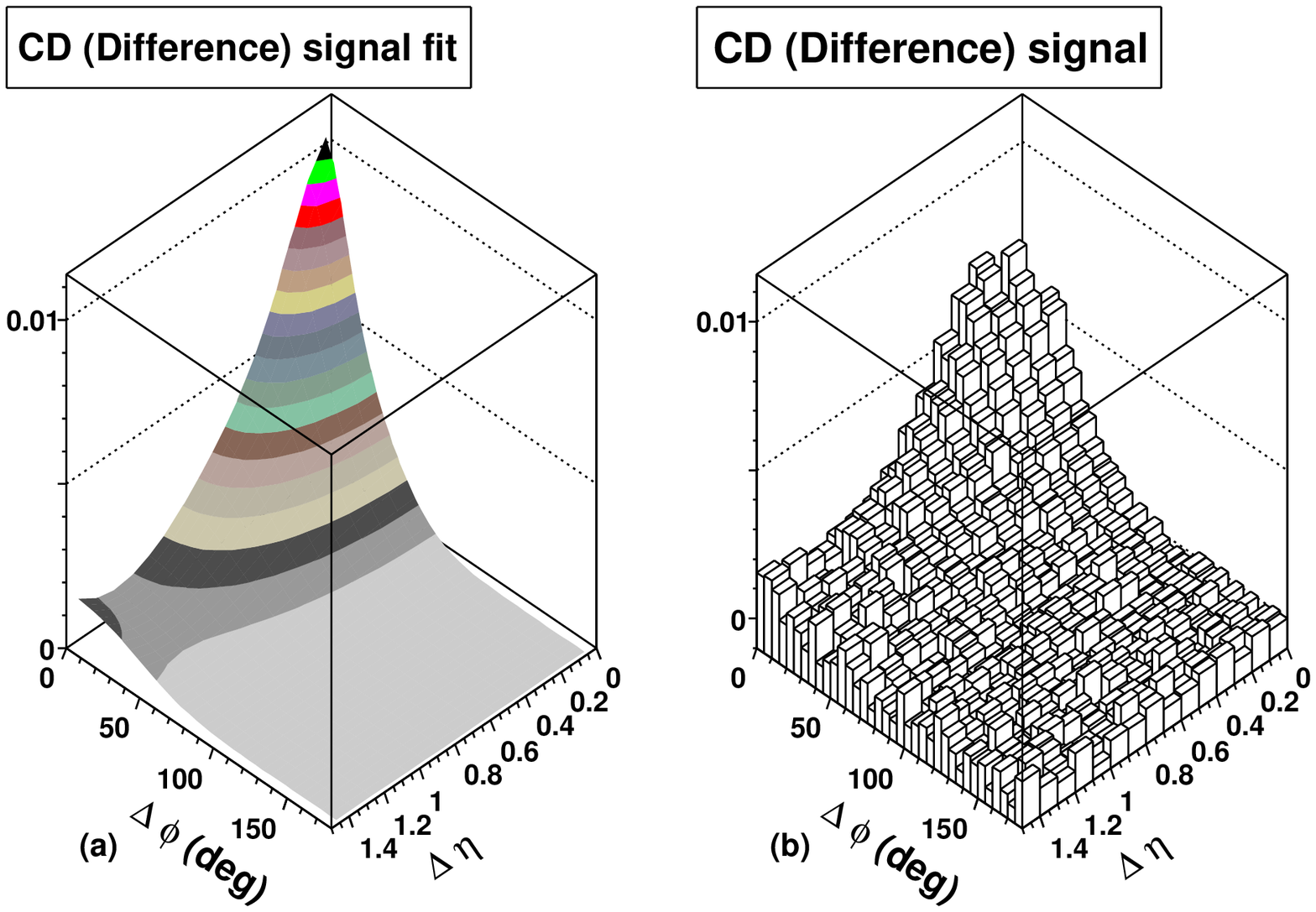}} \caption[]{``(Color online)''a) left side: The charge dependent 
(CD) signal shape is a 2 dimensional (2-D) approximate gaussian which is 
symmetrical. The average gaussian width is $27.5^\circ \pm 3.2^\circ$ 
(0.48 $\pm$ 0.056 rad) both in $\Delta \phi$ and $\Delta \eta$. Thus we are 
observing a greater probability for unlike-sign charge pairs of particles than 
for like-sign charge pairs emitted randomly on 2-D $\eta \phi$ directions.

b) right side: The CD signal data corresponding to the adjacent signal fit on
the left.} 
\label{figure6}
\end{figure*}

The complex multi dimensional $\chi^2$ surface makes the $\chi^2$ increase
non linearly with the number of error ranges ($1\sigma$) given in the Tables I 
and II. Therefore, in order to determine the significance of a parameter or 
group of parameters one must fit without them, and determine by how many 
$\sigma$ the fit has worsened. Then one uses the normal distribution curve to 
determine the significance of the omitted parameter(s). The results of such an 
investigation were:

\begin{enumerate}
\item Leave out $cos(3\Delta \phi)$ and the fit worsens by $95\sigma$. 
\item Leave out dip and the fit worsens by $28\sigma$.
\item Leave out Sector terms and the fit worsens by $11.7\sigma$.
\item Leave out $\eta$ dependent $cos(\Delta \phi)$ and $cos(3\Delta \phi)$ 
terms and the fit worsens by $4.3 \sigma$.
\item Leave out $\eta$bump terms and the fit worsens by $1.54\sigma$.
\item Leave out sector 2 and the fit worsens by $1.41\sigma$.
\end{enumerate}

Since our global fit with all the above terms is at the $3\sigma$ level,
we conclude that the above terms are necessary. The huge number of $\sigma$
we get for some of the parameters(s) are only to be taken as a qualitative 
indication of the need for the parameter(s) to fit our high precision data.
Due to the uncertainties in determining the number of $\sigma$ far out on the 
tails of the normal distribution (i.e. $> 10\sigma$), a quantitative 
interpretation of them cannot be made.

\section{CD and CI Signals}

\subsection{Charge Dependent (CD) Signal} 
If we subtract the entire like-sign charge pairs correlation (equation (5)) 
from the unlike-sign charge pairs correlation (equation (4)) we obtain the CD 
correlation. However it is observed that the background \bf (Bk) \rm of the 
two terms are close enough in value to cancel each other in the subtraction. 
Thus the CD signal is essentially the same as the entire CD correlation. In the
extensively studied balance function \cite{balfun} for the correlation due to
unlike-sign charge pairs which are emitted from the same space and time 
region, it is argued that the emission correlation of these pairs can be 
approximately estimated from the balance function. The balance function for 
these charged pairs is proportional to the unlike-sign charge pairs minus the
like-sign charge pairs. Therefore, the CD signal which is approximately equal
to the CD correlation is a qualitative measure of the emission correlation of 
unlike-sign charge pairs emitted from the same space and time region. In 
addition to the approximations involved in the balance function, there is
the modification to the CD signal discussed in Section IV D.

 Fig. 6a shows the fit to the CD signal. Fig. 6b shows the CD signal data 
that was fit. The signal form is a simple gaussian in both $\Delta \phi$ and 
$\Delta \eta$. The gaussian width in $\Delta \phi$ is 
$28.3^\circ \pm 3.4^\circ$ (0.49 $\pm$ 0.059 rad) and the gaussian 
$\Delta \eta$ width is 0.485 $\pm$ 0.054.  Converting this pseudo-rapidity to 
a $\theta$ angle yields a width of $26.7^\circ \pm 3.0^\circ$ 
(0.47 $\pm$ 0.052 rad). This correlation has the same angular range in 
$\Delta \eta$ and $\Delta \phi$. Therefore we observe pairs of opposite 
charged particles emitted randomly in the $\eta$ and $\phi$ direction with a 
correlation which has an average gaussian width of about $27.5^\circ 
\pm 3.2^\circ$ (0.48 $\pm$ 0.056 rad) in the $\theta$ angle corresponding 
to $\Delta \eta$ and also in the azimuth $\Delta \phi$. 

\begin{figure*}[ht] \centerline{\includegraphics[width=0.800\textwidth]
{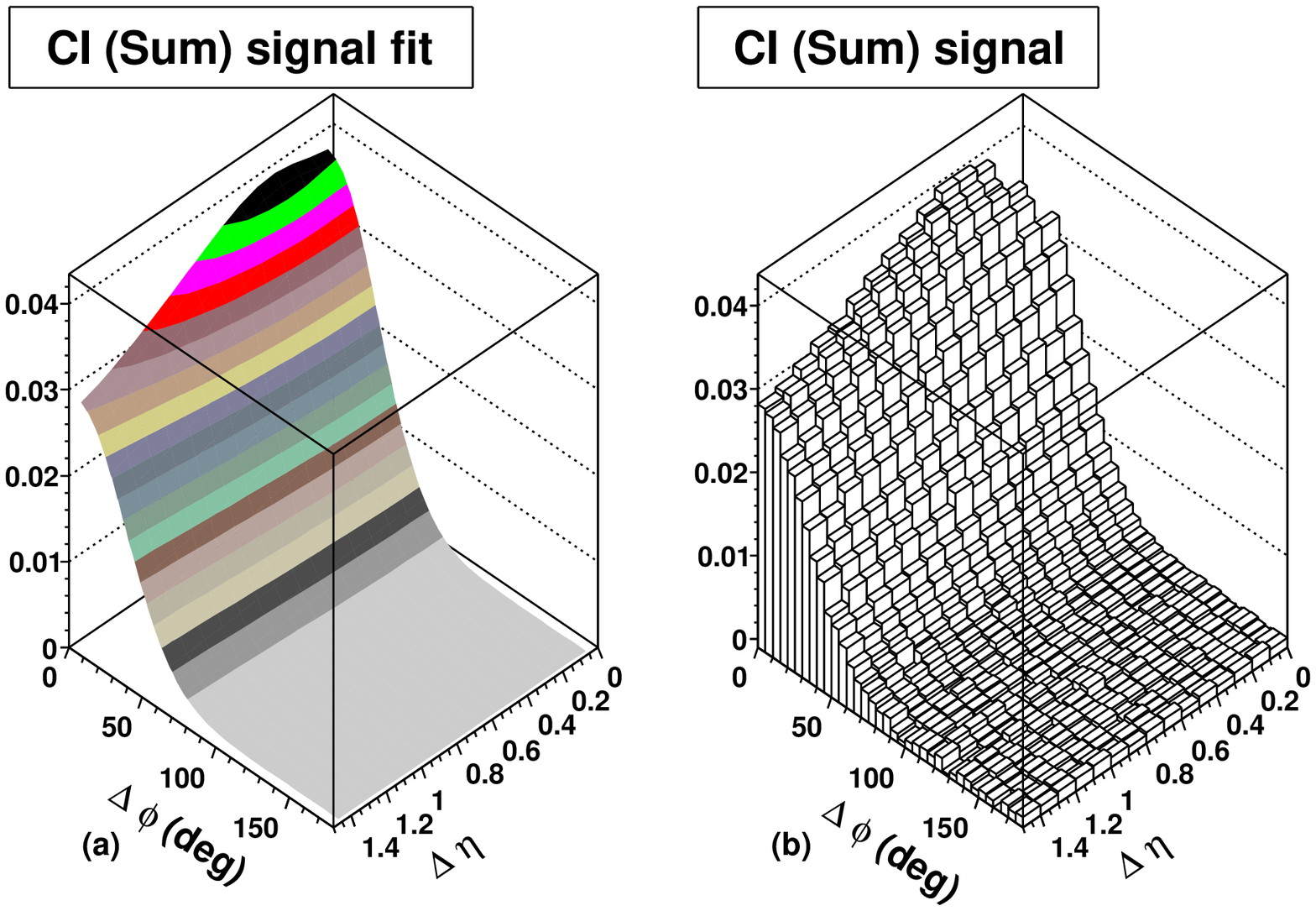}} \caption[]{``(Color online)''a) left side: The fit to the charge
independent (CI) signal shape which is the sum of the fits of like-sign plus 
unlike-sign charge pairs signals. The 2-D gaussian equivalent RMS signal has a 
$\Delta \phi$ width of about $32^\circ$ (0.52 rad), and $\Delta \eta$ width of 
about $66^\circ$ (1.15 rad) which is about double the $\Delta \phi$ width.

b) right side: The CI signal data corresponding to the 
adjacent fit on the left.} 
\label{figure7}
\end{figure*}

\subsection{Charge Independent (CI) Signal}
 If we add the like-sign charge pairs signal to the unlike-sign charge pairs
signal, we obtain the CI signal. The CI signal fit shown in Fig. 7a 
displays the average structure of the correlated emitting sources. 
Fig. 7b shows the CI signal data which was fit by the analytic distribution 
shown in Fig. 7a.

In order to get a good fit to the shape, both in $\Delta \eta$ and 
$\Delta \phi$, we needed a more complicated form than one simple 2-D gaussian. 
The CI signal actually contains an approximate gaussian from the unlike-sign 
charge pairs signal plus two gaussians from the like-sign charge pairs signal.

We want to obtain a measure of the effective $\Delta \phi$ and $\Delta \eta$
widths of the overall pattern of the CI signal. A good method for doing this 
is to compare the CI signal with a single gaussian which yields the same root 
mean square (RMS) values as the actual good fit described above involving two 
gaussians and an approximate gaussian. For $\Delta \phi$ that $\sigma$ is 
$32.0^\circ \pm 0.6^\circ$ (0.56 $\pm$ 0.01 rad).  For $\Delta \eta$ that 
$\sigma$ is 1.55 corresponding to an angle of 
$66.0^{\circ+1.0^\circ}_{-0.6^\circ}$ (1.15 $\pm$ 0.02 rad). Thus the 
correlation is about twice as wide in $\Delta \theta$, the angle corresponding 
to $\Delta \eta$, than in the angle corresponding to $\Delta \phi$ 
(see Fig. 7a).

Another STAR measurement reports CD and CI correlations at $\sqrt{s_{NN}}$ = 
130 GeV/c \cite{aya}.\footnote{The CD correlation in Ref. \cite{aya} is defined
as like-sign pairs minus unlike-sign charge pairs which has opposite algebraic 
sign relative to the definition used in this paper}. The major differences 
between that previous analysis and the present work are the larger range in 
$p_t$ $0.15 < p_t  < 2.0$ GeV/c and the lower statistical quality of the 
limited 130 GeV/c dataset. Although the analysis in Ref. \cite{aya} included 
low $p_t$ particles there is reasonable qualitative agreement with the present 
results.

\subsection{Resonance Contribution}   
To determine the maximum contribution and effect resonances can have on the CI
signal we consider the following. Resonances mainly decay into two particles.
Therefore, neutral resonances which decay into charged pairs could be a 
partial source of the unlike-sign charge pairs in CD and CI signals. However,
resonances do not add significant correlation to the like-sign charged pairs in
either the CD or the CI signals. 

We expect that the final state particles from the resonance decay distribution 
will be symmetrical in the $\Delta \phi$ angle and the angle corresponding to
$\Delta \eta$ since we do not expect any polarization mechanism that would 
disturb this expected symmetry. The fact that in the CD both the $\Delta \phi$
angle and the angle corresponding to $\Delta \eta$ are approximately the same
supports this. In this section we are evaluating the possible effects of 
resonance decay on the CI signal which definitely is observed to be elongated 
in the corresponding $\Delta \eta$ angle by about a factor of 2 compared to the
$\Delta \phi$ angle. The CI signal is always defined as: CI signal = 
unlike-sign charge pairs signal + like-sign charge pairs signal. CD signal =  
unlike-sign charge pairs signal - like-sign charge pairs signal. By subtracting
the CD signal from the CI signal and rearranging terms in the equation one 
obtains the following.  CI signal = CD signal + 2$\cdot$(like-sign charge pairs
signal). 

There is particular interest in the shape of the CI signal when comparisons are
made to theoretical models e.g. HIJING and Ref. \cite{themodel}. The 
$\Delta \phi$ shape is about the same in the CI and CD signals. Therefore, we 
need only estimate the effect of resonances on the $\Delta \eta$ width. By 
making the extremely unrealistic assumption that the CD signal is composed 
entirely (100\%) of resonances, we can estimate the maximum effect of 
resonances on the CI signal $\Delta \eta$ width. We use the equivalent gaussian
which has the same root mean square (RMS) widths for this calculation. From the
data we determine that the $\Delta \eta$ width of the CI signal is increased by
a 7\% upper limit. However if we use the result from the Appendix B that the 
resonance content of the CD is 20\% or less, we must divide the 7\% by 5 
which results in an increase of $\Delta \eta$ width of only about 2\%. Either 
estimate of $\Delta \eta$ width increase is inconsequential, compared to the 
observed approximate factor of two elongation of the $\Delta \eta$ width 
compared to the $\Delta \phi$ width. Details of these calculations are given 
in Section VI Systematic Errors and Appendix B.

\begin{figure*}[ht] \centerline{\includegraphics[width=0.800\textwidth]
{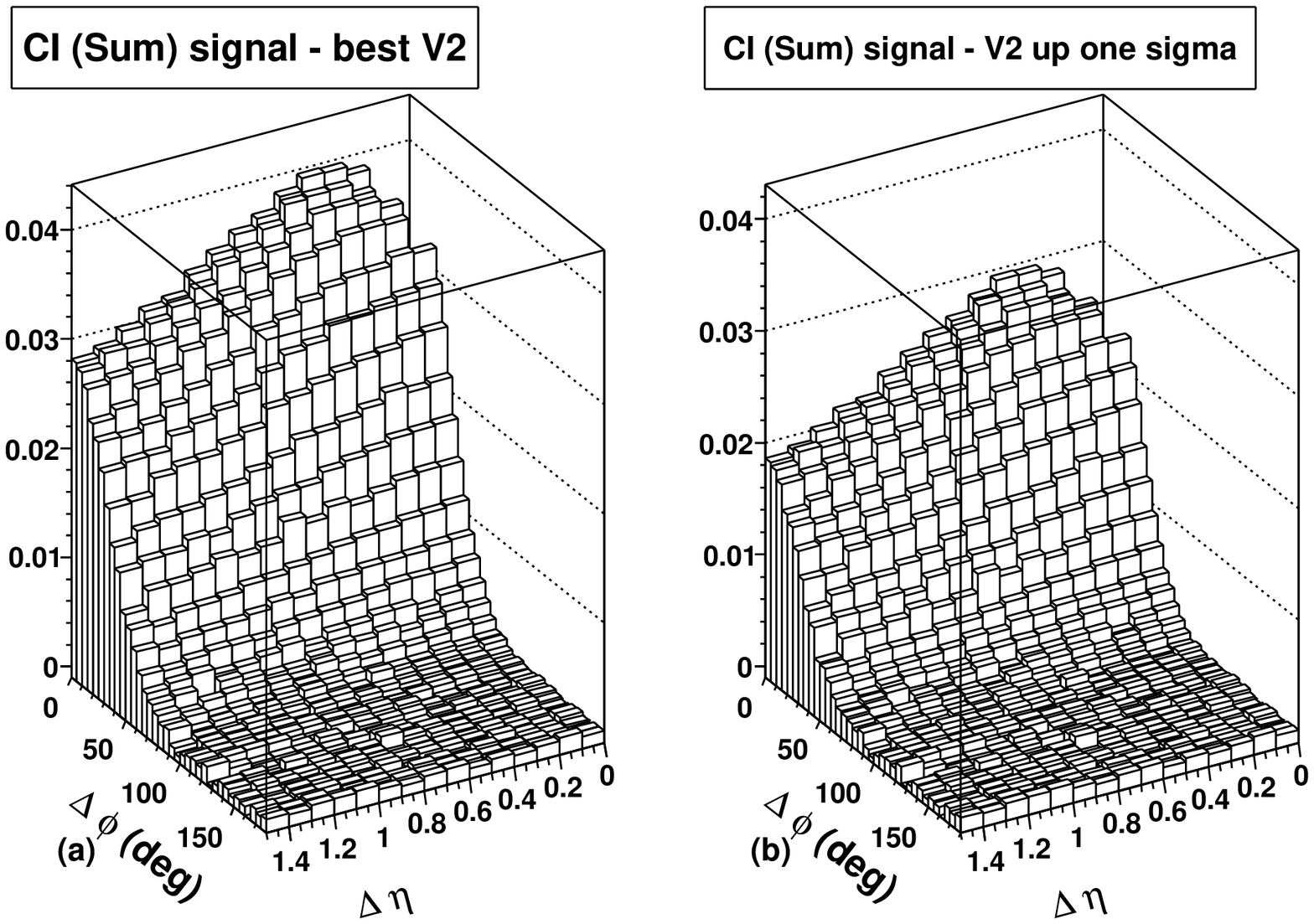}} \caption[]{a) left side: The charge independent (CI) signal data 
with the lower range of elliptic flow amplitude and the best $\chi^2$ (the 
same as Figure 7b) plotted as a 2-D perspective  plot on $\Delta \phi$ vs. 
$\Delta \eta$.

b) right side: The charge independent (CI) signal data with our
maximum elliptic flow amplitude used in our analysis ($\chi^2$ was worse by 
$1\sigma$).} 
\label{figure8}
\end{figure*}

\subsection{Modification of CI and CD Signals}

The CI and CD signals existing at the time of hadronization are changed by the 
continuing interaction of the particles until kinetic freeze-out, when 
interactions cease. The interactions are expected to reduce the signals.
Therefore we expect that the observed signals are less than those existing 
at the time of hadronization.

\section{Net Charge Fluctuation Suppression}

Net charge fluctuation suppression is an observed percentage reduction in the
RMS width of the distribution of the number of positive tracks minus the 
negative tracks plotted for each event, compared to the RMS width of a random
distribution.

If there are localized, uncharged bubbles of predominantly gluons in a color 
singlet, when the bubble hadronizes the total charge coming from the bubbles 
is very close to zero.  Therefore if we are detecting an appreciable sample of 
such bubbles, we expect to see net charge fluctuation suppression.

It should be noted that net charge fluctuation suppression can be deduced
from the CD correlation given previously.  Net charge fluctuation suppression
previously analyzed at lower momenta has been found consistent with resonance 
decay \cite{claude}. However the present analysis has different
characteristics. Furthermore Ref.\cite{themodel} has made specific estimates
for net charge fluctuation suppression applicable to this experiment. We 
performed a charge difference analysis for tracks within cuts of 
$0.8 < p_t  < 2.0$ GeV/c and $| \eta| <  0.75$. This would allow a comparison
with Ref. \cite{themodel}.

For each event we determined the difference of the positive tracks minus
the negative tracks in our cuts. There was a net mean positive charge of 
4.68 $\pm$ 0.009.  The width of the charge difference distribution given by 
(RMS) was $11.149\pm 0.017$. To determine the net charge fluctuation 
suppression we need to compare this width with the width of the appropriate 
random distribution, which would have no net charge fluctuation suppression. 
However we must arrange a slight bias toward a positive charge so that we end 
up with the same net mean positive charge as observed. If we now assign a 
random charge to each track with a slight bias toward being positive such that 
the mean net charge is also $4.68 \pm 0.009$, the width (RMS) becomes 11.865 
$\pm$ 0.017. The percentage difference in the widths which measures the net 
charge fluctuation suppression is $6.0\% \pm 0.2\%$. 

\section{Systematic Errors} 

Systematic errors were minimized using cuts and corrections. The cuts 
(see Section II C) were large enough to make contributions from track 
merging, Coulomb, and HBT effects negligible.

Systematic checks utilized $\chi^2$ analyses which verified that the 
experimental results did not depend on the magnetic field direction, 
the vertex z coordinate, or the folding procedures (see Sections II A and B).
By cutting the track $\eta$ at 1.0 we keep the systematic errors of the track
angles below about $1^\circ$ \cite{backbegone}. 

In Section IV C we referred to a simulation in Ref. \cite{themodel}, which 
estimated that the background resonance contribution to the CD correlation 
cannot be more than 20\% (see Appendix B). From the resonance calculations
discussed in Appendix B the background resonance contribution to the 
unlike-sign charge pairs correlation would be reduced to 5\%. The details that 
justify this are: Our analyses measured an average of 19 unlike-sign charge 
signal pairs per event from the CD correlation signal. The total average number
of unlike-sign charge signal pairs is 88. Therefore with the extreme assumption
 that all of the CD correlation signal is due to background resonances, 21.6\% 
of the unlike-sign charge signal pairs are due to resonances. However as 
discussed in Appendix B the estimated contribution of resonances to the CD 
correlation is 20\% or less. Therefore 21.6\%/5 = 4.3\% is the resonance 
content of the unlike-sign charge signal pairs, which we rounded to 5\%. 

We also concluded via an upper limit calculation with the above extremely 
unrealistic assumption that the shape of the CI signal could only have the 
measured width in $\Delta \eta$ increased by about 7\%. This is inconsequential
compared to the factor of two increase in $\Delta \eta$ width compared to 
$\Delta \phi$ width. For the simulation discussed above, in which resonance 
backgrounds contribute 20\% to the CD correlation, the increase in the CI 
correlation $\Delta \eta$ width would be limited to about 2\%.
 
The CI signal and the CD signal are more physically significant than the
unlike-sign and like-sign charge pairs signals, since they are physically
interpretable. The CI signal fit displays the average structure of the
correlated emitting sources (Section IV B). The analysis suggests that the CD 
signal (Section IV A) is a qualitative representation of the emission 
correlation of the unlike-sign charge pairs emitted from the same space and 
time region. Elliptic flow contributions to the like and unlike-sign charge 
pairs signals are approximately equal and therefore cancel in the CD signal.
Thus the CD signal is not affected by uncertainty in the elliptic flow 
amplitude.

In Fig. 8a we show the CI signal data with the lower flow range value of mean 
elliptic flow amplitude (weighted over the $p_t$ range of the data) used in
the fit, namely, $v_2$ = 0.035. This is our best fit, in which the value of 
$v_2$ is consistent with the lower flow range. Thus Fig. 8a is the same as 
Fig. 7b. In Fig. 8b we show the CI signal data with our maximum amount of 
elliptic flow allowed. This value of elliptic flow causes our fit to be 
$1\sigma$ worse, and corresponds to a $\chi^2$ increase of 32. This results in 
a mean weighted $v_2$ = 0.042. This value of elliptic flow lies in the 
determined flow range, but is smaller than the upper limit of 0.047 
\cite{flow}. The upper error ranges in Tables I and II are determined by the 
effect of this change in the elliptic flow (0.042$> v_2 >$ 0.035), since it is 
the dominant error which results in these range values. The main change in Fig.
8b compared to Fig. 8a is that the peak amplitude is reduced by 25\%, and the 2
dimensional area is reduced by 34\%. However the most significant CI signal 
parameters in comparing to theoretical models are those that measure the shape,
namely the ratio of RMS $\Delta \phi$ and $\Delta \eta$ widths, of a gaussian 
equivalent to the fit. For our best fit $\Delta \phi$ width = $32.0^\circ \pm 
0.6^\circ$ (0.56 $\pm$ 0.01 rad), and $\Delta \eta$ width = 1.55 which is 
equivalent to $66.0^{\circ+1.0^\circ}_{-0.6^\circ}$ (1.15 $\pm$ 0.02 rad). The 
ratio $\Delta \eta$ equivalent angle/ $\Delta \phi$ = 2.06. For the case of the
maximum elliptic flow value the determined widths were $\Delta \phi$ = 
$30.1^\circ$ (0.53 rad), and $\Delta \eta$ = 1.375 ($61.6^\circ$or 1.08 rad). 
The ratio $\Delta \eta$ / $\Delta \phi$ = 2.05. Thus the important shape ratio 
has changed about 1\%. 

Let us now address the errors due to contamination by including secondary
particles arising from weak decays and the interaction of anti-protons and
other particles in the beam pipe and material near the beam pipe. These
secondary particle backgrounds have been estimated to be about 10-15\% 
\cite{starback}. In this analysis we are concerned mainly with the angles of 
the secondary particles  relative to the primaries that survive our high $p_t$ 
cut, not their identity or exact momentum magnitude. Our correlations almost 
entirely depend on angular measurements of $\Delta \phi$ and $\Delta \eta$. 
In the range $0.8 < p_t < 2.0$ GeV/c we have considered the behavior of 
weakly decaying particles and other non-primary particles which could 
satisfy our distance of closest approach to the primary vertex and $p_t$ cuts. 
Because high $p_t$ secondary particles are focused in the same direction as 
the primaries, only a fraction of these particles have sufficient change in 
angle that would cause an appreciable error in our correlation. Hence our 
estimate is a systematic error of about 4\% due to secondary particles.

Below we summarize our extensive discussion of systematic errors in this 
section. Track merging errors, Coulomb, and HBT effects were made negligible
by cutting out effected bins at small space angles. Instrumental errors were 
corrected for in the parameterization. Elliptic flow error is our dominant 
systematic error, due to the uncertainty that exists in the elliptic flow
analyses \cite{flow}. It should be noted that the size of the upper error range
of our parameter errors in Table I and II are determined by the maximum value 
of the elliptic flow we allowed ($v_2$=0.042) which corresponds to a $\chi^2$ 
change of 32 ($1\sigma$). The sizeable change to the CI signal shown in Fig. 
8b compared to Fig. 8a is due to this maximum allowed value of elliptic flow.
As shown above fortunately the shape ratio of the CI signal which is important 
for theoretical comparison is changed only by about 1\%. When more accurate 
elliptic flow results become available it is a simple matter to insert these 
in the fits and reduce the resultant error. The next smaller systematic error 
is due to possible background resonance errors. The smallest systematic error 
is due to secondary contamination. As discussed above none of these errors 
affect our important physical conclusions significantly.  

\section{Discussion} 

Highly significant correlations are observed for unlike-sign charge pairs and
like-sign charge pairs, and consequently for charge dependent (CD) and charge
independent (CI) signals. The CD signal is well described (with a $1\sigma$ 
fit) by a symmetrical 2-D gaussian with an RMS width of about $30^\circ$ in 
both $\Delta \phi$ and $\Delta \eta$. Conservative simulations in Ref. 
\cite{themodel} indicate that the contribution to this CD signal from 
background resonance decay is less than 20\%. Simulations (see Appendix B) 
estimate that the only significant change due to the background resonances is
an increase in the CD amplitude of 20\% or less. The CI signal is more complex 
and is the sum of the unlike-sign charge pairs signal and the like-sign charge 
pairs signal fits. Therefore it contains an approximate gaussian from the 
unlike-sign charge pairs signal, and a large positive gaussian plus a small 
negative gaussian (dip) from the like-sign charge pairs signal. The dip 
contains only 1.6\% of the like-sign charge signal volume. This small dip, 
observed for the first time, has high significance in the fit. This feature is 
consistent with what one would expect for suppression of like-sign charge pair 
emission from a localized neutral source such as gluons.

A 2-D gaussian which yields the same RMS widths as our overall CI signal,
has a $\sigma$ along the $\Delta \phi$ direction of $32.0^\circ \pm 0.6^\circ$ 
(0.56 $\pm$ 0.01 rad).
However the $\Delta \eta$ $\sigma$ is 1.55 corresponding to an angle of
approximately $66^\circ$ (1.15 rad) and thus is twice as wide.

The mean charge difference for tracks within the chosen analysis cuts is 
4.68 $\pm$ 0.009 net positive charges, and the RMS variation of this quantity 
from event to event is 11.149 $\pm$ 0.017. A random charge assignment 
constrained to produce the same mean net charge has a larger width of 11.865
$\pm$ 0.017. The difference, $6.0\% \pm 0.2\%$, measures the net charge 
fluctuation suppression. 

The HIJING model produces jets in our $p_t$ range which are nearly symmetrical 
in $\Delta \phi$ and the angle corresponding to $\Delta \eta$ \cite{Wang}. 
The proper way to compare our data with HIJING is to compare our CI signal 
with the above mentioned HIJING CI correlation from jets. As shown in IV B, 
our CI signal is highly asymmetric since the angle corresponding to 
$\Delta \eta$ is about twice the angle $\Delta \phi$ thus strongly 
contradicting these basic characteristics of HIJING jets correlations.

Ref.\cite{PBM} presents a detailed comparison of the STAR data presented here 
with a model based on a ring of localized bubbles emitting charged particles 
from the central fireball surface at kinetic freeze out. Good consistency with 
the STAR data presented in the present paper is reported\cite{PBM}.  

\section{Summary and Conclusions} 

We performed an experimental investigation of particle-pair correlations in 
$\Delta \phi$ and $\Delta \eta$ using the main Time Projection Chamber of the 
STAR detector at RHIC. We investigated central Au + Au collisions at 
$\sqrt{s_{NN}}$ = 200 GeV, selecting tracks having transverse momenta 
$0.8 < p_t < 2.0$ GeV/c, and the central pseudo-rapidity region 
$|\eta| < 1.0$. The data sample consists of 2 million events, and the 
symmetries of the data in $\Delta \eta$ and $\Delta \phi$ allow four quadrants 
to be folded into one. The entire data set (unlike-sign charge pairs and 
like-sign charge pairs) was fit by a reasonably interpretable set 
of parameters, 17 for the unlike-sign charge pairs and 19 for the like-sign
charge pairs. These parameters are small in number compared to the total number
of degrees of freedom, which was over 500 for each of the two types of pairs. 
Every fit reported here using these parameters was a good fit of 2 to 3 
$\sigma$ or less.

Section VI discusses systematic errors. From our analysis of the systematic 
errors we conclude that the important features of our data and the conclusions 
drawn have not been significantly affected by the systematic errors.

Section VII discusses the results and model fits.

This paper serves as an excellent vehicle for making detailed comparisons with,
and testing of various theoretical models such as the Bubble Model 
\cite{themodel} and other relevant models. 
 
\section{Acknowledgment} 

We thank the RHIC Operations Group and RCF at BNL, and the
NERSC Center at LBNL for their support. This work was supported
in part by the Offices of NP and HEP within the U.S. DOE Office
of Science; the U.S. NSF; the BMBF of Germany; IN2P3, RA, RPL, and
EMN of France; EPSRC of the United Kingdom; FAPESP of Brazil;
the Russian Ministry of Science and Technology; the Ministry of
Education and the NNSFC of China; IRP and GA of the Czech Republic,
FOM of the Netherlands, DAE, DST, and CSIR of the Government
of India; Swiss NSF; the Polish State Committee for Scientific
Research; SRDA of Slovakia, and the Korea Sci. \& Eng. Foundation.

\appendix

\section{Multi-parameter fitting in the large DOF region and systematic
uncertainties}

Let us now consider the detail of multi-parameter data analysis in the large
DOF region in terms of change of $\chi^2$ for our 517-519 n for unlike-sign
and like-sign charge pairs respectively. A $1\sigma$ change in the
significance of the individual fits require the change in $\chi^2$ of 32.
The reader is referred to Ref.\cite{probability} from which we quote:
``For large n(DOF), the $\chi^2$ p.d.f.(probability density function)
approaches a gaussian with mean = n and variance($\sigma$) squared = 2n.''
For n$>$50-100 this result has been considered applicable, and it remains
applicable and becomes more accurate as n increases toward infinity. Thus
for our 517-519 n, $1\sigma$ = $\sqrt{2n}$ =32.

The statistical significance of any fit in this paper can be obtained by the 
following procedure: The number of $\sigma$'s of fit = ($\chi^2$ - n)/32.
The number of $\sigma$'s refer to the normal distribution curve.

For large DOF(bins - parameters) fluctuations occur because of the many bins.
When one fits the parameters, they will try and describe some of these 
fluctuations. Therefore we need to check whether the fluctuations in the data 
sample are large enough to significantly distort the parameter values. This has
traditionally been done by using the confidence level tables vs. $\chi^2$ which
allows a reasonable determination of the fluctuations due to binning. The
approximation (described above) we have used is an accurate extension of the
confidence level tables.

Let us consider our method of assigning systematic error ranges to the 
parameters. Our objective is to obtain error ranges which are not likely to be 
exceeded if future independent data samples  taken under similar conditions are
obtained by repetition of the experiment by STAR or others. We want to avoid 
the confusion and uncertainty of apparent significant differences in parameters
when the fit to new data does not differ significantly from a previous fit. We 
allow each parameter, one at a time, to be varied (increased and then 
decreased) in both directions while all the other parameters are free to 
readjust until the overall fit $\chi^2$ degrades in significance by $1\sigma$. 
This corresponds to an increase of $\chi^2$ of 32 for the unlike-sign and 
like-sign charged pairs fits. The $\chi^2$ surface has been observed, and 
$\chi^2$ increases very non linearly with small increases of the parameter 
beyond the error range.

When one assumes that the parameter error is given by a $\chi^2$ change of 1
in the best fit\cite{statistics}, this is correct for the case where you know
the underlying physics. The purpose of our use of parameters is not to 
determine their true values, since we do not know the underlying physics. We 
use the parameters to determine an analytic description of the data that fits 
it within 2-3$\sigma$.

Let us address how one determines the significance of a parameter or a 
particular group of parameters in the case of this multi-parameter analysis. 
It is entirely incorrect to compare the error ranges in Table I and II with the
parameter value. The only correct way to determine the significance of a
parameter or group of parameters in the multi-parameter fit is to leave the 
parameter(s) out of the fit and refit without the parameter(s). One then
determines by how many $\sigma$'s the overall fit has degraded compared to
a change of 32 for $1\sigma$.

One should note that the upper error range on a parameter in Tables I and II
represents the change in value of the parameter as we increase $v_2$, our
dominant error, from our best fit until the fit worsens by $1\sigma$ ($\chi^2$
increases by 32). The other parameters are free to readjust during the 
foregoing variation. The lower error range is the result of changing the 
parameter in the opposite direction of it in the upper range while the other
parameters are free to readjust until the fit decreases in significance by 
$1\sigma$. The largest error which dominates the upper systematic error range 
determination is the variation of the $v_2$ error. However the constraint of 
$v_2$ to the flow range, and the fact that our best fit is at the lower end of 
the range means that flow is not varied when we determine the lower error.
Therefore the elimination of the large flow variation leads to generally 
smaller lower error ranges and thus observed asymmetry between the upper and 
lower error ranges.

Finally if one fixes any one of the parameters at the end of the error ranges 
with the correct $v_2$ assigned (0.042 upper and 0.035 lower) and lets all the 
other parameters readjust, one will achieve an overall fit which is only worse 
by $1\sigma$ (a $\chi^2$ increase of 32).
 
\section{resonance calculations}
    
In this section we discuss background resonances which have the following
meaning. Background resonances are resonances that only contribute to the 
unlike-sign charge pair correlation. Any resonance that is emitted as part of
a correlated emitter will also have a like-sign charge correlation. This is
because any of the charges will have another particle of the same charge to 
correlate with which is also emitted by the correlated emitter. The background
resonances we are discussing are not part of the correlated emitters we want 
to study, but will add to the CD correlation.

In Ref. \cite{themodel} a table of resonances and particles was considered
in a thermal resonance gas model. The resonance content was adjusted to cause
the 20\% net charge fluctuation suppression observed at lower momentum
in a STAR experiment for central Au + Au collisions. The result of this 
model calculation is a 20\% contribution of resonances to the CD 
correlation. The $p_t$ range of the analysis $0.8 < p_t < 2.0$ GeV/c tends to 
have the symmetric pairs of particles from resonance production decay into 
angles within the range of our cuts. However asymmetric decays only contribute 
one particle of the pair and thus do not add to the correlation. Therefore 
effectively only a part of the estimated resonance contribution affects the 
correlation measurement. Our simulations reveal that the only significant 
change that the background resonance contribution would introduce in the CD 
correlation is an increase in amplitude of 20\% or less. The CI signal = CD 
signal + 2$\cdot$(like-sign charge pairs signal). The resonance contribution 
to the like-sign charge pairs signal is small. The affect of resonances on the 
CI signal elongation in the $\Delta \eta$ width is estimated by the model 
simulation to be a factor five lower than the extreme maximum case previously 
assumed, that the entire CD correlation is composed of background resonances. 
Thus this would result in a $\Delta \eta$ width elongation of the CI signal of 
approximately 2\% instead of the upper limit of 7\%.

\end{document}